# Observation of doubly-degenerate topological flatbands of edge states in strained graphene

Yongsheng Liang[1†], Jingyan Zhan[1†], Shiqi Xia[1*], Daohong Song[1,2*], Zhigang Chen[1,2*]

[1]*The MOE Key Laboratory of Weak-Light Nonlinear Photonics, TEDA Applied Physics Institute and School of Physics, Nankai University, Tianjin 300457, China*

[2]*Collaborative Innovation Center of Extreme Optics, Shanxi University, Taiyuan, Shanxi 030006, China*

[†]*These authors contributed equally to this work*

*Corresponding author: shiqixia@nankai.edu.cn, songdaohong@nankai.edu.cn, zgchen@nankai.edu.cn*

**Abstract:**

Flat bands are of significant interest due to their potential for energy confinement and their ability to enable strongly correlated physics. Incorporating topology into flatband systems further enhances flatband mode robustness against perturbations. Here, we present the first realization of doubly degenerate topological flatbands of edge states in chiral-symmetric strained graphene. The flatband degeneracy stems from Dirac point merging, achieved by tuning the coupling ratios in a honeycomb lattice with twig boundary conditions. The nontrivial topology of these modes is characterized by the winding of the Berry connection, which ensures their robustness against disorder. Experimentally, two types of topological edge states are observed in a strained photonic graphene lattice, consistent with numerical simulations. Moreover, the degeneracy of the topological flatbands doubles the density of states for zero-energy modes, facilitating the formation of compact edge states and enhancing control over edge states and light confinement. Our findings underscore the interplay among lattice geometry, symmetry, and topology in shaping doubly degenerate topological flatbands. This opens new possibilities for advancements in correlated effects, nonlinear optical phenomena, and efficient energy transfer in materials science, photonic crystals, and quantum devices.

Flatband systems are characterized by their dispersionless energy spectra and vanishing group velocity [1, 2], features that enable the emergence of strongly correlated phenomena such as the fractional quantum Hall effect, Mott insulating states, and unconventional superconductivity [3-7]. These unique properties arise from the macroscopic degeneracy of states at a given energy level, enhancing interaction effects and enabling exotic quantum phases. A variety of fundamental phenomena, including Anderson localization [8-10], Landau–Zener Bloch oscillations [11], and compact localized states [12-17], have been explored to deepen the understanding of flatband physics. However, the susceptibility of flatband states to disorder and perturbations presents a significant challenge, limiting their applications. In contrast, topological states, which are protected by global topological invariants, exhibit enhanced robustness and transport properties, even in the presence of disorder [18-21]. This inherent robustness has already been leveraged in various advanced applications, including topological lasers [22-26], terahertz transmission [27-29], and quantum photonics [30-32], where stability and coherence are critical. Recent experiments have demonstrated that the synergy between flatbands and topological phases can be realized through engineered edges in honeycomb lattices [33] or through embedded flux [34, 35], greatly enhancing the robustness of flatband modes. Nevertheless, the degeneracy of topological flatbands remains unexplored, which could offer immense potential for advancements in areas such as slow light [36, 37], enhanced light-matter interaction [38], and dispersionless image transmission [39, 40]. Achieving flatband degeneracy with desired features, however, requires precise control over both flatband conditions and topological invariants, posing a significant challenge.

Graphene, a single layer of carbon atoms arranged in a honeycomb lattice (HCL), has long been considered as a paradigm of materials due to its remarkable electronic properties [41]. Apart from carbon-based real materials, engineered HCL structures have been realized in optics [42-51], with photonic graphene emerging as a versatile platform to emulate the electronic behavior of graphene, particularly the edge-related effects. The four fundamental boundary conditions—bearded, armchair, zigzag, and twig—have been proposed and experimentally demonstrated using photonic graphene [33, 43, 45]. Notably, graphene edge states, characterized by nontrivial winding number [52], can be found under the bearded, zigzag, and twig boundary conditions, where they are degenerate and exhibit flat dispersion relations. Therefore, by tailoring boundary configurations and introducing external modulations, graphene offers a promising platform to investigate the interplay between flatband physics, topology, and degeneracy.

In this work, we unveil the existence of doubly degenerate topological flatbands of edge states in chiral-symmetric strained graphene. The first flatband arises from the twig boundary in the HCL with uniform coupling (referred to here as "unstrained graphene"). The second flatband emerges by applying a uniaxial strain to the HCL along a specific direction, thereby tuning the couplings (referred to as "strained graphene"). Both flatbands consist of topological edge states and are characterized by nontrivial winding of lattice Hamiltonian. This flatband degeneracy leads to doubling of the density of states (DOS) for topological zero-energy modes, enabling the formation of two distinct sets of elongated edge states along with two sets of compact localized edge states (CESs) at the boundary. Experimentally, we observe such edge states in photonic graphene with the twig boundary condition (TBC), where coupling ratios (controlled by the relative positions between sites) are tuned by judiciously applying a strain on the HCL. Supported by numerical simulations, the existence of doubly degenerate topological flatbands is conclusively verified. These findings reveal new possibilities for enhancing DOS of topological states, which are useful, for example, in flatband lasing.

We begin by considering an HCL with the TBC, as depicted in Fig.1(b). Under the tight-binding approximation, there are two sites (black and white filled circles) in each unit cell and the nearest-neighbor couplings are denoted as $t_a$ and $t_b$ (black and orange lines in Fig.1(b)). The bulk Hamiltonian can be expressed as:

$$H(\mathbf{k}) = \begin{pmatrix} 0 & h(\mathbf{k}) \\ h^*(\mathbf{k}) & 0 \end{pmatrix} \quad (1)$$

with $h(\mathbf{k}) = t_a + t_b e^{i\mathbf{k}\mathbf{a_1}} + t_a e^{i\mathbf{k}\mathbf{a_2}}$, where $\mathbf{a_1}$ and $\mathbf{a_2}$ are the basis vectors of the unit cell (highlighted by purple rhombus in Fig.1(b)). In the case of an unstrained HCL, three couplings are equal and typically set to $t_0$. The twig boundary supports edge states (edge states I) with zero-energy degeneracy, forming a complete flatband that spans the entire one-dimensional Brillouin zone (1D BZ) [33], as shown by the green line in Fig. 1(a1). In contrast to 2D Chern insulators, which are characterized by the Chern number, the nontrivial topology of these degenerate edge states aligns with that of the SSH model [53]. It originates from the bulk topological properties and is characterized by the winding number, expressed as:

$$w = \frac{1}{2\pi} \oint \frac{d}{d\mathbf{k}} \arg[h(\mathbf{k})] \, d\mathbf{k} \quad (2)$$

where $h(\mathbf{k})$ is the off-diagonal term of the bulk Hamiltonian $H(\mathbf{k})$. For the twig boundary along the x-direction (Fig.1(b)), the integral is performed along a closed loop in $k_y$-direction for a fixed $k_x$. The winding loop at the boundary of the 1D BZ ($k_x = \pi/\sqrt{3}a$) is plotted in the ($\sigma_x, \sigma_y$) plane in Fig. 1(a2). As $k_y$ completes one period, the loop completes two circles. However, only the larger circle encircles the origin (denoted by an origin dot) in ($\sigma_x, \sigma_y$) plane, resulting in $w = 1$ and the presence of edge states I. In contrast, for the other three boundary conditions (zigzag, bearded, and armchair), the winding can form only a single loop in the ($\sigma_x, \sigma_y$) plane. In other words, in an unstrained HCL with any edge termination, the winding number is always restricted to either $w = 1$ or $w = 0$, but not larger. Further details are provided in Supplementary Material [54]. Since the twig boundary condition offers the possibility for doubling the winding, one may wonder if it is possible to achieve doubly degenerate topological states by enabling the winding to encircle the origin twice (i.e., to realize $w = 2$). We show that this is indeed possible and can be realized by tuning the coupling ratio $\delta$, defined as $\delta = t_b/t_a$ (Fig. 1(b)), i.e., in a strained HCL. When a uniaxial strain (compression) is applied along the coupling direction of $t_b$, the couplings become highly non-uniform, resulting in $\delta > 1$ and leading to changes in both the bulk and edge properties. The DOS for the ribbon under TBC is shown in Fig. 1(c) for different $\delta$ values, with the peaks corresponding to the zero-energy flatband of edge states, and the gray region representing the bulk bands. For $\delta < 2$, the DOS of zero-energy modes increases significantly, indicating the emergence of new edge states (edge states II) that degenerate with edge states I at the zero energy. For $\delta > 2$, the bulk band gap opens, and the DOS of zero-energy modes remains constant but has a doubled value as compared to that for the unstrained HCL ($\delta = 1$). During this process, the smaller circle of the winding loop under TBC begins to grow larger, and eventually, both circles encircle the origin (Fig. 1(d2)), thus yielding $w = 2$ at any given $k_x$. The nontrivial winding implies the existence of edge states II across the entire 1D BZ, forming a new flatband (red line in Fig. 1(d1)) after the gap opens, which becomes fully degenerate with edge states I.

To directly illustrate the formation of the doubly degenerate topological flatbands, the distributions of the vector field of $h(\mathbf{k})$ under different coupling ratios are shown in Fig. 2. For the two-band HCL with chiral symmetry, the integral of the vector field along the $k_y$-path determines the winding number at a fixed $k_x$. For the unstrained HCL ($\delta = 1$), the inequivalent Dirac points (red and blue dots in Fig. 2(a2)) reside at the corners of the first BZ (orange hexagon), and the path taken by the integral follows the green vertical line in Fig. 2(a). The green shaded region corresponds to the first 1D BZ for the twig boundary. It is evident that the green arrows of the vector $h(\mathbf{k})$ make a complete loop in this region, resulting in $w = 1$. According to the principle of "bulk-edge correspondence", the twig edge states (edge states I) appear across the entire 1D BZ, with their energy degenerate at zero due to chiral symmetry, thus forming a topological flatband (green lines in Fig. 2(a1)).

When $\delta > 1$, the Dirac points move along the boundary of the 2D BZ, as indicated by the red and blue arrows in Fig. 2(b2). This shift creates a red-shaded region where the vector $h(\mathbf{k})$ winds twice, resulting in $w = 2$. Consequently, the winding-2 implies the presence of an additional set of topological edge states (edge states II) in the red-shaded region, as shown by the red line in the projected 1D band structure (Fig. 2(b1)). Importantly, tuning the coupling ratio by straining of an HCL preserves its chiral symmetry, keeping edge states II pinned at zero energy while edge states I remain intact. As the coupling ratio $\delta$ reaches the critical value ($\delta = 2$), the Dirac points merge at the M-point (inset in Fig. 2(c2)). Further increasing of $\delta$ results in the gap opening (Fig. 2(c1)), leading to a semimetal-to-insulator transition. This transition establishes the winding-2 region across the entire 1D BZ (red-shaded region in Fig. 2c), where the nontrivial topology at each momentum $k_x$ features the flatband edge modes. The energies of the edge modes are pinned at zero due to the preserved chiral symmetry, leading to the formation of doubly degenerate topological flatbands of edge states (green and red lines in Fig. 2(c1)). Furthermore, flatbands with more than two-fold degeneracy can be achieved by introducing long-range A-B sublattice couplings [54]. We note that, different from the chiral edge modes in Chern insulators, here the bulk structure preserves the time-reversal symmetry, and the edge states forming the flatbands exhibit topological properties originating from the SSH topology. The flatness and degeneracy of the flatband are protected by chiral symmetry, ensuring robustness against any perturbations that preserve this symmetry [54]. The degeneracy not only doubles the DOS at zero energy but also supports two distinct sets of topological edge states at any $k_x$. The two edge states at the boundary of the 1D BZ ($k_x = \pi/\sqrt{3}a$) are shown in Fig. 2(d1, d2), and the edge states at other $k_x$ are provided in Supplementary Material [54]. Although both edge states are localized on the same sublattice (black circles) due to chiral symmetry, they differ in their phase distributions: edge state I exhibits opposite phases at the outermost A sublattices (marked by 1 and 3) of each supercell, whereas edge state II exhibits identical phase. Additionally, the doubly degenerate topological flatbands give rise to two distinct sets of CESs. It is important to note that this double degeneracy of topological flatbands can only be achieved under TBC in the HCL (Fig. 1(b)). Tuning the coupling ratios under other boundary conditions, such as zigzag, bearded, and armchair edges, results in either a single flatband or a completely destroyed flatband when the strain is introduced [43, 55]. The presence of a single flatband or doubly degenerate flatbands crossing the entire first 1D BZ in strain graphene can also be confirmed by the relation between edge states and supercells [54].

In the experiment, we demonstrate the existence of the degenerate topological flatbands in photonic graphene by observing the edge states at different $k_x$. As analyzed above, the topological flatband associated with edge states I is always present, whereas the fully flatband of edge states II emerges only after the gap opens at $k_x = \pi/\sqrt{3}a$ ($\delta > 2$). Here, we focus on the demonstration of the evolution of edge states II, while the results for edge states I are provided in Supplementary Material [54]. The photonic HCL with TBC is fabricated using the continuous-wave laser-writing technique in a nonlinear crystal (SBN) [56]. By applying a strain (compression) to the photonic lattice, the distance between lattice sites is altered, thereby tuning the coupling ratio. Experimentally, two HCLs with different coupling ratios ($\delta = 1.5$ and $\delta = 3$) are established. An example of the strained HCL with $\delta = 3$ is shown in Fig. 3(a), where the nearest-neighbor distances are $d_1 = d_2 = 40.5 \mu m$ and $d_3 = 30.4 \mu m$, and the uniaxial strain direction is indicated by the gray arrow. The probe beams with specific amplitude and phase distributions, matching the theoretically calculated edge modes [54], are generated using a spatial light modulator and sent into the strained HCL. Under different coupling ratios, the probe beams are configured with two different transverse momenta to match the eigenmode distributions of edge states II at $k_x = 0$ and $k_x = \pi/\sqrt{3}a$, all exhibiting identical phase at the outermost A sublattices (marked by 1 and 3) of each supercell (Fig. 3(b1, b2, c1, c2)). The Fourier spectra and phase distributions of the input beams are shown in the insets. The corresponding outputs after $20\ mm$ propagation exhibit distinct difference (Fig. 3(b, c)). Specifically, at $\delta = 1.5$, the probe beam corresponding to edge state II remains localized at $k_x = 0$ (Fig. 3(b1)) but couples into neighboring sites at $k_x = \pi/\sqrt{3}a$ (Fig. 3(b2)), indicating that the structure cannot support a fully flatband of edge states II before the gap opens. After the gap opens ($\delta = 3$), the probe beam remains localized at the edge and in one sublattice for both $k_x = 0$ and $k_x = \pi/\sqrt{3}a$ after

20 $mm$ of propagation (Fig. 3(c1, c2)), suggesting the formation of fully flatband. For comparison, in-phase mixed bulk modes are also considered, where light spreads into the bulk after propagation under both strain conditions of $\delta = 1.5$ and $\delta = 3$. Simulation results for longer propagation distance are provided in Supplementary Material [54]. These results confirm the formation of edge states II throughout the 1D BZ once the gap opens.

In addition to the two sets of elongated edge states, two compact localized edge states (CES I and CES II) arising from the doubly degenerate flatbands are also supported in the strained HCL. The real-space distributions of the two distinct CESs are shown in Fig. 4(a1) (CES I) and Fig. 4(b1) (CES II), with each holding the characteristic phase distribution of their corresponding edge states. CES I exhibits opposite phases at the outermost A sublattices, while CES II exhibits in-phase characteristics (insets in Fig. 4(a1) and Fig. 4(b1)). In momentum space, the occupation of CESs at different momenta is quantified as $\eta(k_x) = |\langle \varphi_{k_x} | \Psi \rangle|^2$, where $\Psi$ represents mode of CES I or CES II, and $\varphi_{k_x}$ is the eigenstate of all elongated edge states at $k_x$. The occupation of CES I and CES II across the entire 1D BZ (Fig. 4(a2, b2)) results in their compact form along the boundaries in real space. Experimentally, probe beams matching the CES I and CES II are launched into the lattice to demonstrate their compactness. Both probe beams remain intact after 20 $mm$ of propagation (Fig. 4(a3, b3)). In contrast, the in-phase probe beams fail to sustain compactness, with light spreading into the B sublattices (Fig. 4(a4, b4)). Furthermore, the degeneracy of all edge states enables any linear combination of the two CESs to serve as eigenmodes of the structure. Here we show a linear combination CES I + CES II, where the energy is nearly localized at a single site at the boundary (Fig. 4(c1)). This highly compact edge state remains localized at the initially excited A sublattice sites after propagation (Fig. 4(c3)), whereas under in-phase excitation, it becomes distorted and spreads into the neighboring B sublattices (Fig. 4(c4)). These experimental results, along with corresponding simulations [54], confirm the formation of doubly degenerate flatbands. The degeneracy and superposition of the two CESs enable precise tuning of the intensity distribution of the compact edge states.

In conclusion, we have demonstrated the existence of doubly degenerate topological flatbands in strained photonic graphene under the twig boundary condition. In contrast to the zigzag, bearded, and armchair boundary conditions, only the twig boundary condition facilitates the unique double degeneracy of flatband edge states, which in turn doubles the DOS of edge modes. The topology of flatbands is characterized by nontrivial double winding in momentum space. The topological flatbands give rise to two distinct sets of robust edge states and CESs, which are experimentally observed in our laser-written photonic graphene. Experimental results agree well with numerical simulations. Our work highlights the interplay between lattice geometry, symmetry, and topology, presenting a new approach for controlling topological flatbands. This framework opens exciting possibilities for advancements in non-Abelian physics [51-53], topological lasing [54], and robust energy confinement [55-57]. By leveraging the doubly degenerate flatband structure, our findings suggest new methods for enhancing light-matter interactions, enabling the design of compact and efficient photonic devices and expanding the frontier of topological phases in engineered materials.

## Acknowledgement

This work was supported by National Key R&D Program of China (No. 2022YFA1404800); the National Nature Science Foundation of China (No. 12134006, 12274242, and 12474387); the Natural Science Foundation of Tianjin (No. 21JCJQJC00050) and the 111 Project (No. B23045) in China.

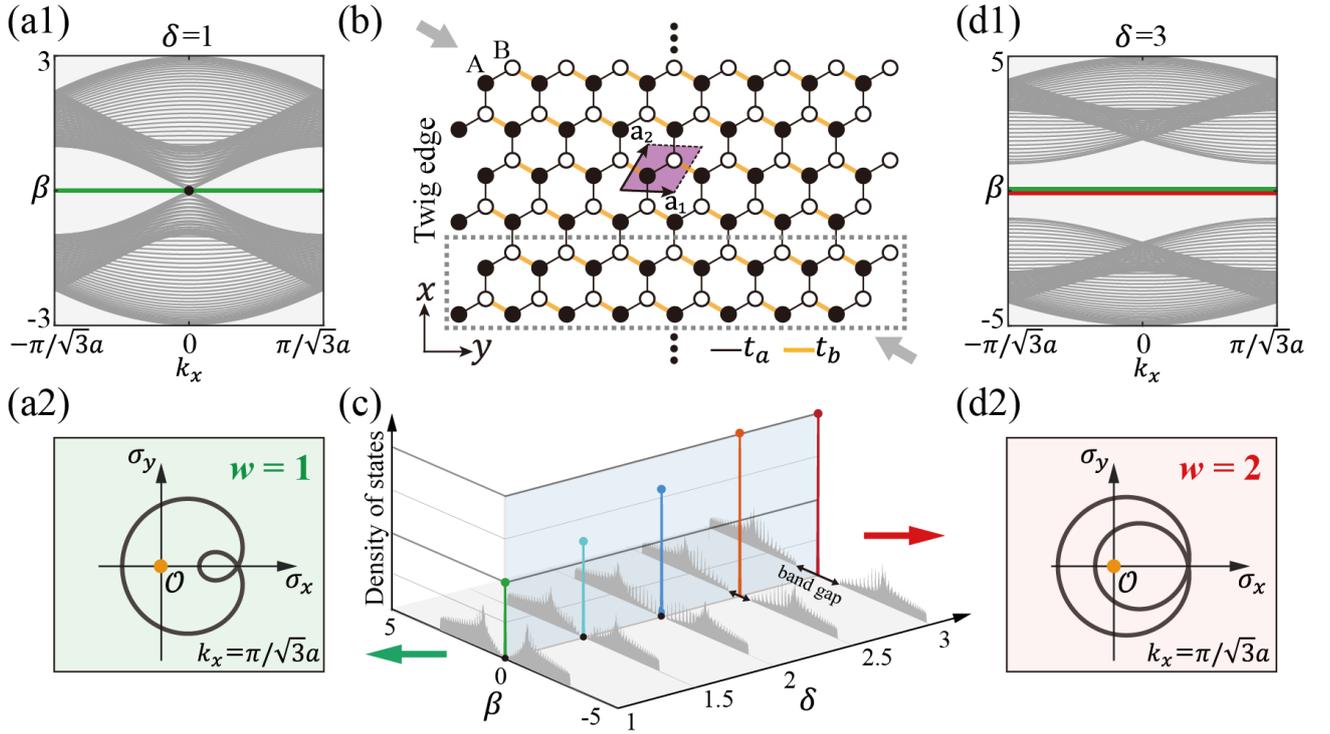

**Fig.1 Doubly degenerate topological flatbands in a strained HCL with twig boundaries.** (a1) The 1D band structure of an unstrained HCL ($\delta = t_b/t_a = 1$) with twig boundaries, where the green lines represent the region of edge states. (a2) Winding loop for the twig boundary in the ($\sigma_x$, $\sigma_y$) plane at $k_x = \pi/\sqrt{3}a$. The orange dot marks the origin $\mathcal{O}$. (b) Schematic diagram of the HCL with twig boundaries and periodic along the $x$-direction. The purple-shaded rhombus marks the unit cell with two sublattices ($A$ and $B$), and the gray dashed rectangle indicates the supercell corresponding to the twig boundary. The primitive vectors are $\mathbf{a_1} = a\hat{y}$ and $\mathbf{a_2} = \sqrt{3}a/2\hat{x} + a/2\hat{y}$, where a is the lattice constant. Black and orange lines represent the weak couplings ($t_a$) and strong couplings ($t_b$) of the strained HCL, respectively, with the strain applied along the $t_b$ coupling direction indicated by two gray arrows. (c) Evolution of the density of states (DOS) as a function of the coupling ratio ($\delta$). The DOS of the topological zero modes is highlighted by colors. (d1, d2) Plots have the same layout as (a1, a2) but for a strained HCL ($\delta = 3$), displaying two degenerate flatbands pinned to zero energy (green and red lines in (d1)). In this latter case, the DOS of zero modes as well as the nontrivial winding number is doubled compared to (a).

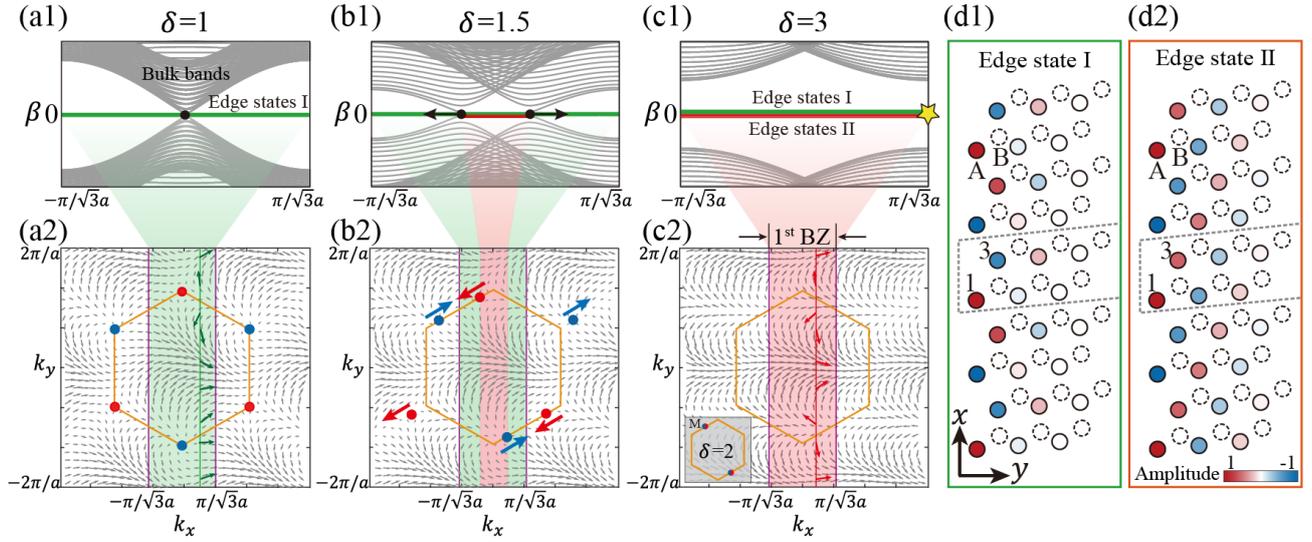

**Fig.2 The topological origin of doubly degenerate flatbands in a strained HCL.** (a1-c1) 1D band structure of the HCL with twig boundaries under different coupling ratios at $\delta = 1$ (a1), $\delta = 1.5$ (b1), and $\delta = 3$ (c1). The black dots mark the degenerate points, and the black arrows in (b1) indicate the movement of degenerate points under compression. The original twig edge states (edge states I) are highlighted by green lines, and the new edge states (edge states II) are shown in red lines. (a2-c2) The corresponding schematics for the winding number calculation, where the arrows point in varying directions of the vector $\mathbf{h}(\mathbf{k})$. Red and blue dots represent Dirac points with opposite Berry phases. The winding number takes the value of $w = 1$ ($w = 2$) in the green (red) shaded region, which determines the number of edge states. The inset in the bottom-left corner of (c2) shows the merging of Dirac points at the $M$ point when $\delta = 2$. (d) Illustration of the edge state distributions corresponding to edge states I (d1) and edge states II (d2) at $k_x = \pi/\sqrt{3}a$, denoted by the yellow star in (c1). Red and blue dots in (d) represent opposite phase distributions.

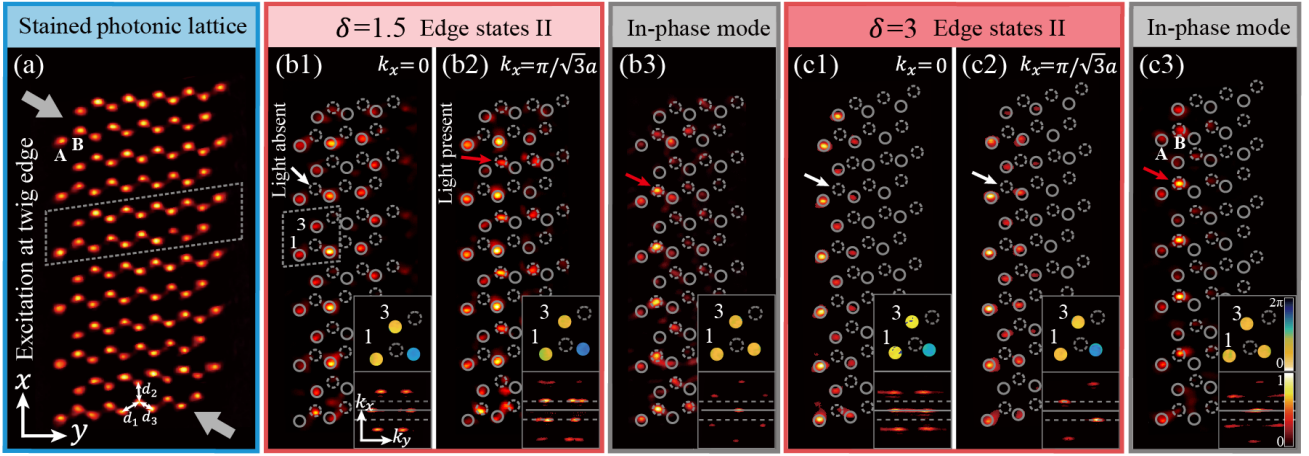

**Fig.3 Experimental observation of edge states II in strained photonic graphene.** (a) A laser-written strained graphene lattice with twig boundary along the $x$-direction and the strain applied along $d_3$ direction at $\delta = 3$. In this case, the distances between nearest-neighbor sites are $d_1 = d_2 = 40.5 \mu m$ and $d_3 = 30.4 \mu m$. For a different strain at $\delta = 1.5$, $d_1 = d_2 = 40.5 \mu m$ and $d_3 = 34.5 \mu m$. The gray arrows indicate the uniaxial strain direction. $A$ and $B$ are two sublattices within the unit cell. (b) Experimental outputs of the probe beam matching the eigenmode of edge states II at $k_x = 0$ (b1) and $k_x = \pi/\sqrt{3}a$ (b2) under $\delta = 1.5$. (b3) Experimental outputs of in-phase probe beam. The corresponding phase distributions within the dashed squares and Fourier spectra of the input beam are shown in the insets. The solid (dashed) lines in the Fourier spectra mark the center (edge) of the 1D BZ. (c) Results presented in the same layout as (b) but they are for edge states II under $\delta = 3$. The red (white) arrows in (b) and (c) indicate the presence (absence) of light on the $B$ sublattices. For all the experimental results, the propagation distance is $20\ mm$.

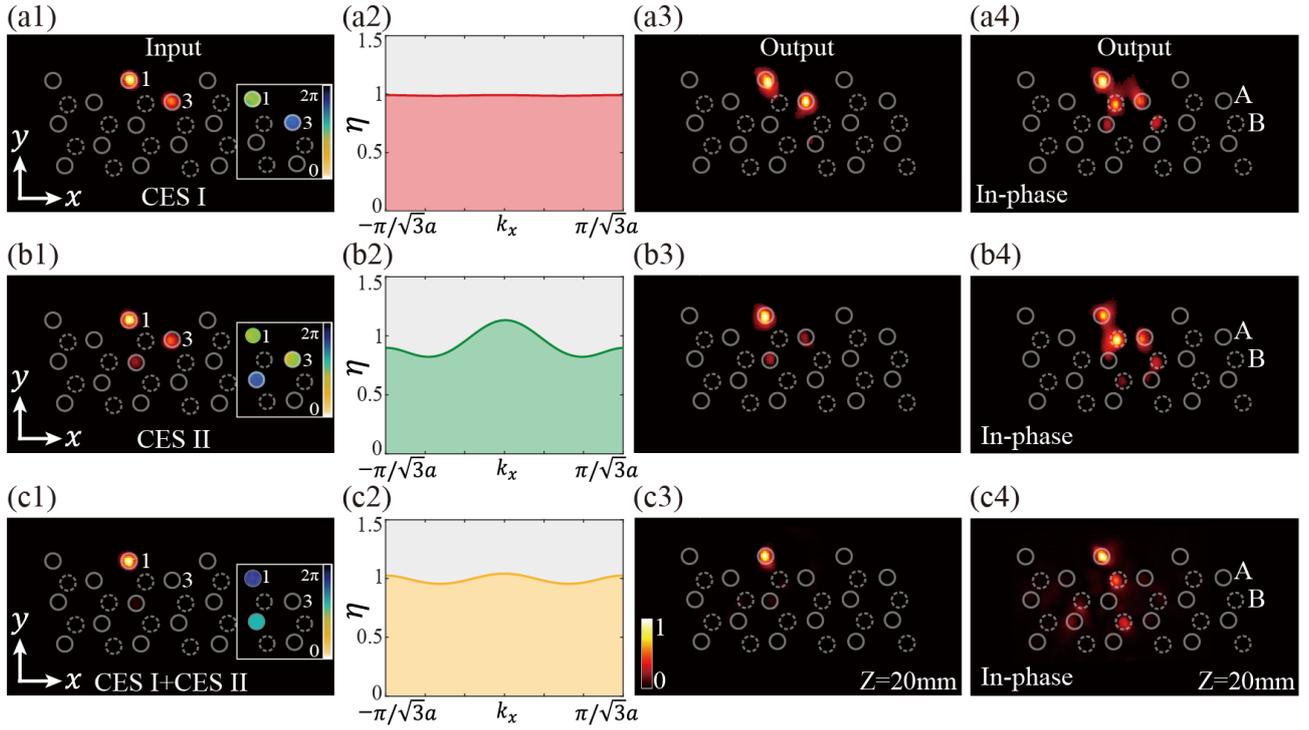

**Fig.4 Demonstration of compact edge states (CESs).** (a1) Intensity and phase (inset on the right) distributions of the probe beam, consistent with the mode distribution of CES I. (a2) Spectrum distribution of CES I in the 1D BZ, where $\eta(k_x)$ is normalized. (a3) Output of the probe beam shown in (a1) after $20\ mm$ of propagation. (a4) Output of the in-phase beam as input for comparison. (b, c) Panels (b1-b4; c1-c4) have the same layout as (a1-a4), but they are for the demonstration of CES II (b) and the linear combination of CES I + CES II (c).

**References:**
1. B. Sutherland, *Localization of electronic wave functions due to local topology.* Phys. Rev. B. **34**, 5208-5211 (1986).
2. D. Leykam, A. Andreanov, and S. Flach, *Artificial flat band systems: from lattice models to experiments.* ADV PHYS-X. **3**, 1473052 (2018).
3. E. Tang, J.-W. Mei, and X.-G. Wen, *High-Temperature Fractional Quantum Hall States.* Phys. Rev. Lett. **106**, 236802 (2011).
4. T. Neupert, L. Santos, C. Chamon, and C. Mudry, *Fractional Quantum Hall States at Zero Magnetic Field.* Phys. Rev. Lett. **106**, 236804 (2011).
5. Y. Cao, V. Fatemi, A. Demir, S. Fang, S. L. Tomarken, J. Y. Luo, J. D. Sanchez-Yamagishi, K. Watanabe, T. Taniguchi, E. Kaxiras, R. C. Ashoori, and P. Jarillo-Herrero, *Correlated insulator behaviour at half-filling in magic-angle graphene superlattices.* Nature. **556**, 80-84 (2018).
6. Y. Cao, V. Fatemi, S. Fang, K. Watanabe, T. Taniguchi, E. Kaxiras, and P. Jarillo-Herrero, *Unconventional superconductivity in magic-angle graphene superlattices.* Nature. **556**, 43-50 (2018).
7. C. Xu and L. Balents, *Topological Superconductivity in Twisted Multilayer Graphene.* Phys. Rev. Lett. **121**, 087001 (2018).
8. M. Goda, S. Nishino, and H. Matsuda, *Inverse Anderson Transition Caused by Flatbands.* Phys. Rev. Lett. **96**, 126401 (2006).
9. T. Schwartz, G. Bartal, S. Fishman, and M. Segev, *Transport and Anderson localization in disordered two-dimensional photonic lattices.* Nature. **446**, 52-5 (2007).
10. J. T. Chalker, T. S. Pickles, and P. Shukla, *Anderson localization in tight-binding models with flat bands.* Phys. Rev. B. **82**, 104209 (2010).
11. R. Khomeriki and S. Flach, *Landau-Zener Bloch Oscillations with Perturbed Flat Bands.* Phys. Rev. Lett. **116**, 245301 (2016).
12. D. L. Bergman, C. Wu, and L. Balents, *Band touching from real-space topology in frustrated hopping models.* Phys. Rev. B. **78**, (2008).
13. R. A. Vicencio, C. Cantillano, L. Morales-Inostroza, B. Real, C. Mejía-Cortés, S. Weimann, A. Szameit, and M. I. Molina, *Observation of Localized States in Lieb Photonic Lattices.* Phys. Rev. Lett. **114**, 245503 (2015).
14. S. Mukherjee, A. Spracklen, D. Choudhury, N. Goldman, P. Öhberg, E. Andersson, and R. R. Thomson, *Observation of a Localized Flat-Band State in a Photonic Lieb Lattice.* Phys. Rev. Lett. **114**, 245504 (2015).
15. S. Xia, Y. Hu, D. Song, Y. Zong, L. Tang, and Z. Chen, *Demonstration of flat-band image transmission in optically induced Lieb photonic lattices.* Opt. Lett. **41**, 1435-1438 (2016).
16. Y. Zong, S. Xia, L. Tang, D. Song, Y. Hu, Y. Pei, J. Su, Y. Li, and Z. Chen, *Observation of localized flat-band states in Kagome photonic lattices.* Opt. Express. **24**, 8877-8885 (2016).
17. J.-W. Rhim and B.-J. Yang, *Classification of flat bands according to the band-crossing singularity of Bloch wave functions.* Phys. Rev. B. **99**, 045107 (2019).
18. M. Z. Hasan and C. L. Kane, *Colloquium: Topological insulators.* Rev. Mod. Phys. **82**, 3045-3067 (2010).
19. X.-L. Qi and S.-C. Zhang, *Topological insulators and superconductors.* Rev. Mod. Phys. **83**, 1057-1110 (2011).
20. F. D. M. Haldane and S. Raghu, *Possible Realization of Directional Optical Waveguides in Photonic*

*Crystals with Broken Time-Reversal Symmetry.* Phys. Rev. Lett. **100**, 013904 (2008).
21. T. Ozawa, H. M. Price, A. Amo, N. Goldman, M. Hafezi, L. Lu, M. C. Rechtsman, D. Schuster, J. Simon, O. Zilberberg, and I. Carusotto, *Topological photonics.* Rev. Mod. Phys. **91**, 015006 (2019).
22. B. Bahari, A. Ndao, F. Vallini, A. El Amili, Y. Fainman, and B. Kanté, *Nonreciprocal lasing in topological cavities of arbitrary geometries.* Science. **358**, 636-640 (2017).
23. M. A. Bandres, S. Wittek, G. Harari, M. Parto, J. Ren, M. Segev, D. N. Christodoulides, and M. Khajavikhan, *Topological insulator laser: Experiments.* Science. **359**, eaar4005 (2018).
24. Y. Zeng, U. Chattopadhyay, B. Zhu, B. Qiang, J. Li, Y. Jin, L. Li, A. G. Davies, E. H. Linfield, B. Zhang, Y. Chong, and Q. J. Wang, *Electrically pumped topological laser with valley edge modes.* Nature. **578**, 246-250 (2020).
25. A. Dikopoltsev, T. H. Harder, E. Lustig, O. A. Egorov, J. Beierlein, A. Wolf, Y. Lumer, M. Emmerling, C. Schneider, S. Höfling, M. Segev, and S. Klembt, *Topological insulator vertical-cavity laser array.* Science. **373**, 1514-1517 (2021).
26. L. Yang, G. Li, X. Gao, and L. Lu, *Topological-cavity surface-emitting laser.* Nat. Photonics. **16**, 279-283 (2022).
27. Y. Yang, Y. Yamagami, X. Yu, P. Pitchappa, J. Webber, B. Zhang, M. Fujita, T. Nagatsuma, and R. Singh, *Terahertz topological photonics for on-chip communication.* Nat. Photonics. **14**, 446-451 (2020).
28. J. Wang, S. Xia, R. Wang, R. Ma, Y. Lu, X. Zhang, D. Song, Q. Wu, R. Morandotti, J. Xu, and Z. Chen, *Topologically tuned terahertz confinement in a nonlinear photonic chip.* Light-Science & Applications. **11**, (2022).
29. W. Wang, Y. J. Tan, T. C. Tan, A. Kumar, P. Pitchappa, P. Szriftgiser, G. Ducournau, and R. Singh, *On-chip topological beamformer for multi-link terahertz 6G to XG wireless.* Nature. **632**, 522-527 (2024).
30. S. Mittal, E. A. Goldschmidt, and M. Hafezi, *A topological source of quantum light.* Nature. **561**, 502-506 (2018).
31. A. Blanco-Redondo, B. Bell, D. Oren, B. J. Eggleton, and M. Segev, *Topological protection of biphoton states.* Science. **362**, 568-571 (2018).
32. T. Dai, Y. Ao, J. Bao, J. Mao, Y. Chi, Z. Fu, Y. You, X. Chen, C. Zhai, B. Tang, Y. Yang, Z. Li, L. Yuan, F. Gao, X. Lin, M. G. Thompson, J. L. O'brien, Y. Li, X. Hu, Q. Gong, and J. Wang, *Topologically protected quantum entanglement emitters.* Nat. Photonics. **16**, 248-257 (2022).
33. S. Xia, Y. Liang, L. Tang, D. Song, J. Xu, and Z. Chen, *Photonic Realization of a Generic Type of Graphene Edge States Exhibiting Topological Flat Band.* Phys. Rev. Lett. **131**, 013804 (2023).
34. G. Cáceres-Aravena, M. Nedić, P. Vildoso, G. Gligorić, J. Petrovic, A. Maluckov, and R. A. Vicencio, *Compact Topological Edge States in Flux-Dressed Graphenelike Photonic Lattices.* Phys. Rev. Lett. **133**, 116304 (2024).
35. Y. Wang, C. Jiang, M. Zhao, D. Zhao, and S. Ke, *Compact topological edge modes through hybrid coupling of orbital angular momentum modes.* Phys. Rev. A. **109**, 063516 (2024).
36. J. Li, T. P. White, L. O'faolain, A. Gomez-Iglesias, and T. F. Krauss, *Systematic design of flat band slow light in photonic crystal waveguides.* Opt. Express. **16**, 6227-6232 (2008).
37. S. Noda, M. Yokoyama, M. Imada, A. Chutinan, and M. Mochizuki, *Polarization Mode Control of Two-Dimensional Photonic Crystal Laser by Unit Cell Structure Design.* Science. **293**, 1123-1125 (2001).
38. Y. Yang, C. Roques-Carmes, S. E. Kooi, H. Tang, J. Beroz, E. Mazur, I. Kaminer, J. D. Joannopoulos,

and M. Soljačić, *Photonic flatband resonances for free-electron radiation.* Nature. **613**, 42-47 (2023).

39. J. Yang, P. Zhang, M. Yoshihara, Y. Hu, and Z. Chen, *Image transmission using stable solitons of arbitrary shapes in photonic lattices.* Opt. Lett. **36**, 772-774 (2011).
40. R. A. Vicencio and C. Mejía-Cortés, *Diffraction-free image transmission in kagome photonic lattices.* Journal of Optics. **16**, 015706 (2014).
41. A. H. Castro Neto, F. Guinea, N. M. R. Peres, K. S. Novoselov, and A. K. Geim, *The electronic properties of graphene.* Rev. Mod. Phys. **81**, 109-162 (2009).
42. O. Peleg, G. Bartal, B. Freedman, O. Manela, M. Segev, and D. N. Christodoulides, *Conical Diffraction and Gap Solitons in Honeycomb Photonic Lattices.* Phys. Rev. Lett. **98**, 103901 (2007).
43. M. C. Rechtsman, Y. Plotnik, J. M. Zeuner, D. Song, Z. Chen, A. Szameit, and M. Segev, *Topological creation and destruction of edge states in photonic graphene.* Phys. Rev. Lett. **111**, 103901 (2013).
44. M. C. Rechtsman, J. M. Zeuner, Y. Plotnik, Y. Lumer, D. Podolsky, F. Dreisow, S. Nolte, M. Segev, and A. Szameit, *Photonic Floquet topological insulators.* Nature. **496**, 196-200 (2013).
45. Y. Plotnik, M. C. Rechtsman, D. Song, M. Heinrich, J. M. Zeuner, S. Nolte, Y. Lumer, N. Malkova, J. Xu, A. Szameit, Z. Chen, and M. Segev, *Observation of unconventional edge states in 'photonic graphene'.* Nat. Mater. **13**, 57-62 (2014).
46. D. Song, V. Paltoglou, S. Liu, Y. Zhu, D. Gallardo, L. Tang, J. Xu, M. Ablowitz, N. K. Efremidis, and Z. Chen, *Unveiling pseudospin and angular momentum in photonic graphene.* Nat. Commun. **6**, 6272 (2015).
47. D. Smirnova, S. Kruk, D. Leykam, E. Melik-Gaykazyan, D.-Y. Choi, and Y. Kivshar, *Third-Harmonic Generation in Photonic Topological Metasurfaces.* Phys. Rev. Lett. **123**, 103901 (2019).
48. O. Jamadi, E. Rozas, G. Salerno, M. Milićević, T. Ozawa, I. Sagnes, A. Lemaître, L. Le Gratiet, A. Harouri, I. Carusotto, J. Bloch, and A. Amo, *Direct observation of photonic Landau levels and helical edge states in strained honeycomb lattices.* Light Sci. Appl. **9**, 144 (2020).
49. M. Bellec, C. Poli, U. Kuhl, F. Mortessagne, and H. Schomerus, *Observation of supersymmetric pseudo-Landau levels in strained microwave graphene.* Light Sci. Appl. **9**, 146 (2020).
50. M. Barsukova, F. Grisé, Z. Zhang, S. Vaidya, J. Guglielmon, M. I. Weinstein, L. He, B. Zhen, R. Mcentaffer, and M. C. Rechtsman, *Direct observation of Landau levels in silicon photonic crystals.* Nat. Photonics. **18**, 580-585 (2024).
51. R. Barczyk, L. Kuipers, and E. Verhagen, *Observation of Landau levels and chiral edge states in photonic crystals through pseudomagnetic fields induced by synthetic strain.* Nat. Photonics. **18**, 574-579 (2024).
52. S. Ryu and Y. Hatsugai, *Topological Origin of Zero-Energy Edge States in Particle-Hole Symmetric Systems.* Phys. Rev. Lett. **89**, 077002 (2002).
53. J. K. Asbóth, L. Oroszlány, and A. Pályi, *A short course on topological insulators.* Vol. 919. 2016: Springer.
54. *See Supplementary Material.*
55. M. Bellec, U. Kuhl, G. Montambaux, and F. Mortessagne, *Manipulation of edge states in microwave artificial graphene.* New J. Phys. **16**, 113023 (2014).
56. S. Xia, A. Ramachandran, S. Xia, D. Li, X. Liu, L. Tang, Y. Hu, D. Song, J. Xu, D. Leykam, S. Flach, and Z. Chen, *Unconventional Flatband Line States in Photonic Lieb Lattices.* Phys. Rev. Lett. **121**, 263902 (2018).